\documentclass[conference,onecolumn]{IEEEtran}

\IEEEoverridecommandlockouts                              
\usepackage{enumerate}

\usepackage{graphicx, epsfig}\usepackage{epstopdf}

\usepackage{algorithmic}
\usepackage{algorithm}

\usepackage{amsmath}
\usepackage{wasysym}

\usepackage{stfloats} 

\usepackage{url}

\usepackage{amsfonts}
\usepackage{amssymb}

\usepackage{multirow}

\makeatletter

\newcommand{\Rmnum}[1]{\expandafter\@slowromancap\romannumeral #1@}
\makeatother

\usepackage{authblk}
\title{\LARGE \bf On Achievable Schemes of Interference Alignment with Double-Layered Symbol Extensions in Interference Channel}
\author{Haichuan Zhou, Tharm Ratnarajah$\dag$\\
 The Institute of Electronics, Communications and Information Technology (ECIT),\\
Queen's University Belfast, Queen's Road, Belfast, UK\\
$\dag$The University of Edinburgh, Edinburgh, UK\\
Email: hzhou01@qub.ac.uk }

\begin{document}

\maketitle \thispagestyle{empty} \pagestyle{empty}

\begin {abstract}
This paper looks into the $K$-user interference channel.
Interference Alignment is much likely to be applied with
double-layered symbol extensions, either for constant channels in
the H$\o$st-Madsen-Nosratinia conjecture or slowly changing
channels. In our work, the core idea relies on double-layered symbol
extensions to artificially construct equivalent time-variant
channels to provide crucial \textit{channel randomness or
relativity} required by conventional Cadambe-Jafar scheme in
time-variant channels \cite{IA-DOF-Kuser-Interference}.
\end {abstract}

\IEEEpeerreviewmaketitle

\section{Introduction}

%


The H$\o$st-Madsen-Nosratinia conjecture was proposed in the
investigation on the multiplexing gain of a network with $K$ source
nodes and $K$ destination nodes in pairs while each node has only a
single antenna and all the nodes could cooperate
\cite{multiplexing-gain-wireless}. The multiplexing gain is also
known as degrees of freedom (DoF) of the channel, in particular
denoting the pre-log factor of the rate as signal-noise-ration (SNR)
approaches infinity. So the question is: how large a multiplexing
gain/DoF is possible for the $K$-pair network. It is proved that for
$K=2$ the DoF is 1. While for the general $K$-pair network, it is
proved that the upperbound of achievable DoF is $K/2$, i.e. the
network obtains at most $K/2$ DoF. However, it is believed that
$K/2$ is not a tight upperbound and furthermore it is conjectured
that the achievable DoF is still only one.

It is important to highlight an implicit condition in the
H$\o$st-Madsen-Nosratinia conjecture, i.e all the channels are
constant. So that it makes difference with the prominent result of
conventional Cadambe-Jafar scheme of interference alignment (IA) as
in \cite{IA-DOF-Kuser-Interference,beamforming-efficient-IA}.
Cadambe-Jafar scheme claims that, on condition the channels are
time-variant, the $K$-pair network is able to approach $K/2$ DoF. If
the channels are constant, Cadambe-Jafar does not work due to its
loss of \textit{channel randomness or relativity} of the network,
which we discuss in detail in following sections.

Later, the H$\o$st-Madsen-Nosratinia conjecture was first settled by
Cadambe, Jafar and Wang as in
\cite{IA-asymmetric-complex-signaling-settling-nosratinia-conjecture-trans}.
It shows that at least 1.2 DoF are achievable on the complex
Gaussian 3-user interference channel with constant coefficients for
almost all values except for a subset of measure zero. A novel idea
of asymmetric complex signaling is proposed, in which the inputs are
chosen to be complex but not circularly symmetric.

In our work, a novel method is proposed
for $K$-user interference network with
constant channels. The rest of this paper is organized as follows.
In Section \Rmnum{2}, the system model and preliminaries are
introduced for the $K$-user network. In Section \Rmnum{3}, the novel
design of double-layered symbol extensions are proposed. In Section
\Rmnum{4}, numerical results and performance are given. In Section
\Rmnum{5}, conclusions and remarks are made.


\section{System Model and Preliminaries}

Define all users in a set $\mathcal{K}=\{1,2,\cdots,K\}$. Denote the
channel from $j$-th source to $k$-th destination as a complex scalar
$\mathrm{h}_{kj}$. Specifically at $t$-th time slot, the channel is
denoted as $\mathrm{h}_{kj}^{[t]}$. 
Use symbol extensions for the design, so define the length of
extensions as $T$, i.e. the dimension of the extended signal vector.
Correspondingly, the effective channel from $j$-th source to $k$-th
destination is denoted as the following equation:
\begin{equation}\label{HMN-diag-H-h}
\mathbf{H}_{kj}=\text{Diag}\{\mathrm{h}_{kj}^{[1]},
\mathrm{h}_{kj}^{[2]}, \cdots, \mathrm{h}_{kj}^{[T]}\}
\end{equation}
in which Diag\{$\cdot$\} represents a diagonal matrix composed of
diagonal elements of the scalar channel coefficients at all $T$ time
slots so that the dimension is set as
$\mathbf{H}_{kj}\in\mathbb{C}^{T\times T}$.

Let each user transmit $d_k$ datastreams, and set each precoder
$\mathbf{V}_{k}\in\mathbb{C}^{T\times d_k}$ respectively. Then the
DoF could be calculated as $(\frac{d_1}{T}, \frac{d_2}{T}, \ldots,
\frac{d_K}{T})$ for all $K$ users and the total DoF of network is
$\frac{(d_1+d_2+\ldots+d_K)}{T}$.

\subsection{Conventional Interference Alignment}

According to Cadambe-Jafar scheme in
\cite{IA-DOF-Kuser-Interference,beamforming-efficient-IA},
interference alignment is implemented by setting the following
condition:

\begin{equation}
\begin{aligned}
  &{\mathbf H}_{1i}{\mathbf V}_{i}={\mathbf H}_{13}{\mathbf V}_{3},\ \ \forall i\in\mathcal{K}\backslash\{1,3\}\\
  &{\mathbf H}_{jk}{\mathbf V}_{k}\prec{\mathbf H}_{j1}{\mathbf V}_{1},\ \ \forall
  j\in\mathcal{K}\backslash\{1,k\},\forall k\in\mathcal{K}\backslash\{1\}
\end{aligned}
\label{HMN-background-conditionKuserI}
\end{equation}

Then the solution of precoders satisfying the condition
(\ref{HMN-background-conditionKuserI}) is given by:

\footnotesize
\begin{equation}\label{HMN-background-K-asymp-1}
  \begin{aligned}
    {\mathbf{V}}_{1}=\left\{\left(\prod_{k,l\in\mathcal{K}\backslash\{1\},k\neq l,(k,l)\neq (2,3)}{(\mathbf{T}_{kl})}^{n_{kl}}\right)
    \cdot\mathbf{w}\Bigg| \forall n_{kl}\leq n\right\}
  \end{aligned}
\end{equation}
\normalsize

\footnotesize
\begin{equation}\label{HMN-background-K-asymp-2}
  \begin{aligned}
    {\mathbf{V}}_{3}=\left\{\mathbf{H}_{21}{(\mathbf{H}_{23})}^{-1}\left(\prod_{k,l\in\mathcal{K}\backslash\{1\},k\neq l,(k,l)\neq
    (2,3)}{(\mathbf{T}_{kl})}^{n_{kl}}\right)
    \cdot\mathbf{w}\right.\\\left.\Bigg| \forall n_{kl}\leq n-1\right\}
  \end{aligned}
\end{equation}
\normalsize

\small
\begin{equation}\label{HMN-background-K-asymp-3}
  \begin{aligned}
    \mathbf{V}_{i}={(\mathbf{H}_{1i})}^{-1}\mathbf{H}_{13}\cdot
    \mathbf{V}_{3}, \quad \forall i\in\mathcal{K}\backslash\{1,3\}
  \end{aligned}
\end{equation}
\normalsize

in which

\small
\begin{equation}\label{HMN-background-T-matrix}
\begin{aligned}
&\mathbf{T}_{kl}=\mathbf{H}_{21}{(\mathbf{H}_{23})}^{-1}\mathbf{H}_{13}{(\mathbf{H}_{k1})}^{-1}\mathbf{H}_{kl}{(\mathbf{H}_{1l})}^{-1}\ \forall l,k\in\mathcal{K}\backslash\{1\}\\
&\mathbf{w}=[1\ 1 \cdots \ 1]^T\in\mathbb{C}^{M\times 1}
\end{aligned}
\end{equation}
\normalsize

\hspace{1mm}

In the solution of (\ref{HMN-background-K-asymp-1}),
(\ref{HMN-background-K-asymp-2}) and
(\ref{HMN-background-K-asymp-3}), the length of symbol extensions is
set as $T={(n+1)}^N+n^N$ where $n\in\mathbb{N}$ and
$N=(K-1)(K-2)-1$. Then the precoders have different dimensions:
$\mathbf{V}_{1}\in\mathbb{C}^{M\times (n+1)^N}$ and
$\mathbf{V}_{j}\in\mathbb{C}^{M\times n^N}, \forall
j\in\mathcal{K}\backslash\{1\}$. So that the first user obtains
$\frac{(n+1)^N}{{(n+1)}^N+n^N}$ DoF and all the other $(K-1)$ users
obtain $\frac{(n)^N}{{(n+1)}^N+n^N}$ DoF for each respectively. When
$n\rightarrow\infty$, the obtained DoF for each user approaches
$1/2$, and the total DoF for the $K$-pair network approaches $K/2$.

\subsection{Constant Channel Issue and Slowly Changing Channel
Issue}


When the channel is constant, then the effective channel of
$\mathbf{H}_{kj}$ in (\ref{HMN-diag-H-h}) becomes:
\begin{equation}\label{HMN-diag-H-h-constant}
\begin{aligned}
&\mathbf{H}_{kj}
=\mathrm{h}_{kj}\mathbf{I}_T\\
&\mathrm{h}_{kj}^{[1]}=\mathrm{h}_{kj}^{[2]}=\cdots=\mathrm{h}_{kj}^{[T]}=\mathrm{h}_{kj}
\end{aligned}
\end{equation}
where $\mathbf{I}_T$ is a $T$-dimensional identity matrix.

Then the intermediate matrix $\mathbf{T}_{kl}$ in
(\ref{HMN-background-T-matrix}) is calculated as
$(\mathrm{h}_{21}\mathrm{h}_{23}^{-1}\mathrm{h}_{13}\mathrm{h}_{k1}^{-1}\mathrm{h}_{kl}\mathrm{h}_{1l}^{-1})\mathbf{I}_T$.
Obviously, the precoders $\mathbf{V}_{1}$ in
(\ref{HMN-background-K-asymp-1}) and $\mathbf{V}_{3}$ in
(\ref{HMN-background-K-asymp-2}) are composed of linear dependent
columns respectively, and so is $\mathbf{V}_{i}$ as in
(\ref{HMN-background-K-asymp-3}). So that the precoding schemes are
not applicable when the channels are constant. The reason is the
$K$-pair network loses \textit{channel randomness or relativity},
which is required for interference alignment as in
\cite{IA-DOF-Kuser-Interference}.

When the channel is slowly changing as in most of realistic
situations, it is necessary to wait for much longer time to combine
all the required number of $T$ time slots to apply IA scheme. The
delay is not tolerable in practice. As a primitive investigation, we
propose a simplified model for slowly changing channels as
following:

\begin{equation}\label{HMN-diag-H-h-slow-changing}
\begin{aligned}
\mathrm{h}_{kj}^{[1]}=\mathrm{h}_{kj}^{[2]}=\cdots=\mathrm{h}_{kj}^{[T/2]}=\mathrm{h}_{kj}^{\star}\\
\mathrm{h}_{kj}^{[T/2+1]}=\mathrm{h}_{kj}^{[T/2+2]}=\cdots=\mathrm{h}_{kj}^{[T]}=\mathrm{h}_{kj}^{\circ}
\end{aligned}
\end{equation}

\section{Proposed Scheme based on Symbol Extensions}

First, observe and analyze the above constant channel issue in
(\ref{HMN-diag-H-h-constant}), and naturally come up with a idea of
artificially fluctuating the symbol extensions of channels to
produce randomness. The procedure is as follows.

\subsection{Unsuccessful Trial: Natural and Naive Fluctuation
Coding}

The natural and naive method is to fluctuate the coding at all nodes
with an additional gain on purpose to construct effective
time-variant channels (not successful although). Let the $j$-th
source node has a gain of $\alpha_j^{[t]}$ at the time slot $t$,
which is randomly drawn from a continuous distribution; and the
$k$-th destination node has a gain of $\beta_k^{[t]}$ at the time
slot $t$, which is also randomly drawn from a continuous
distribution. Then the equivalent channel is denoted as
$\mathrm{\tilde h}_{kj}^{[t]}$ in the following equation:
\normalsize
\begin{equation}\label{HMN-alpha-h-beta}
\begin{aligned}
\mathrm{\tilde
h}_{kj}^{[t]}=\beta_k^{[t]}\mathrm{h}_{kj}^{[t]}\alpha_j^{[t]}
\end{aligned}
\end{equation}
\normalsize

Then observe the time-extended effective channel $\mathbf{H}_{kj}$
in (\ref{HMN-diag-H-h}) becomes:
\begin{equation}\label{HMN-diag-H-h-single-extension-slowly-changing}
\begin{aligned}
\mathbf{H}_{kj}&=\text{Diag}\{\mathrm{\tilde h}_{kj}^{[1]},
\mathrm{\tilde h}_{kj}^{[2]}, \cdots, \mathrm{\tilde
h}_{kj}^{[T]}\}\\
&=\text{Diag}\{\beta_k^{[1]}\mathrm{h}_{kj}^{[1]}\alpha_j^{[1]},
\beta_k^{[2]}\mathrm{h}_{kj}^{[2]}\alpha_j^{[2]}, \cdots, \beta_k^{[T]}\mathrm{h}_{kj}^{[T]}\alpha_j^{[T]}\}\\
\end{aligned}
\end{equation}

It could be further decomposed into a concise and explicit form:
\begin{equation}\label{HMN-diag-H-h-single-extension-slowly-changing-decompose}
\begin{aligned}
\mathbf{H}_{kj}&=\Xi_k\Delta_{kj}\Omega_j\\
\Xi_k&=\text{Diag}\{\beta_k^{[1]},
\beta_k^{[2]}, \cdots, \beta_k^{[T]}\}\\
\Delta_{kj}&=\text{Diag}\{\mathrm{h}_{kj}^{[1]},
\mathrm{h}_{kj}^{[2]}, \cdots, \mathrm{h}_{kj}^{[T]}\}\\
\Omega_j&=\text{Diag}\{\alpha_j^{[1]}, \alpha_j^{[2]}, \cdots,
\alpha_j^{[T]}\}
\end{aligned}
\end{equation}

When the channel is constant, the time-extended effective channel
$\mathbf{H}_{kj}$ in (\ref{HMN-diag-H-h-constant}) becomes:
\begin{equation}\label{HMN-diag-H-h-single-extension}
\begin{aligned}
\mathbf{H}_{kj}&=\text{Diag}\{\mathrm{\tilde h}_{kj}^{[1]},
\mathrm{\tilde h}_{kj}^{[2]}, \cdots, \mathrm{\tilde
h}_{kj}^{[T]}\}\\
&=\mathrm{h}_{kj}\cdot\text{Diag}\{\beta_k^{[1]}\alpha_j^{[1]},
\beta_k^{[2]}\alpha_j^{[2]}, \cdots, \beta_k^{[T]}\alpha_j^{[T]}\}\\
\end{aligned}
\end{equation}

So that in the case of constant channel, observe that
$\mathbf{H}_{kj}$ in (\ref{HMN-diag-H-h-single-extension}) is indeed
an effective time-variant channel. However, to unveil its real
impact for the complete scheme, it is further decomposed with the
following equation:
\begin{equation}\label{HMN-diag-H-h-single-extension-decompose}
\begin{aligned}
\mathbf{H}_{kj}&=\mathrm{h}_{kj}\cdot\Xi_k\Omega_j\\
\Xi_k&=\text{Diag}\{\beta_k^{[1]},
\beta_k^{[2]}, \cdots, \beta_k^{[T]}\}\\
\Omega_j&=\text{Diag}\{\alpha_j^{[1]}, \alpha_j^{[2]}, \cdots,
\alpha_j^{[T]}\}
\end{aligned}
\end{equation}

Then check the intermediate matrix $\mathbf{T}_{kl}$ in
(\ref{HMN-background-T-matrix}) again, with the surprising result as
following: \small
\begin{equation}\label{HMN-background-T-matrix-single-extension}
\begin{aligned}
&\mathbf{T}_{kl}=\mathbf{H}_{21}{(\mathbf{H}_{23})}^{-1}\mathbf{H}_{13}{(\mathbf{H}_{k1})}^{-1}\mathbf{H}_{kl}{(\mathbf{H}_{1l})}^{-1}\\
&=\mathrm{h}_{21}\Xi_2\Omega_1(\mathrm{h}_{23}\Xi_2\Omega_3)^{-1}\mathrm{h}_{13}\Xi_1\Omega_3(\mathrm{h}_{k1}\Xi_k\Omega_1)^{-1}\mathrm{h}_{kl}\Xi_k\Omega_l(\mathrm{h}_{1l}\Xi_1\Omega_l)^{-1}\\
&=\mathrm{h}_{21}(\mathrm{h}_{23})^{-1}\mathrm{h}_{13}(\mathrm{h}_{k1})^{-1}\mathrm{h}_{kl}(\mathrm{h}_{1l})^{-1}\cdot\mathbf{I}_T
\end{aligned}
\end{equation}
\normalsize

Surprisingly in (\ref{HMN-background-T-matrix-single-extension}),
$\mathbf{T}_{kl}$ is still a scaled identity matrix. So that the
precoders $\mathbf{V}_{1}$ in (\ref{HMN-background-K-asymp-1}),
$\mathbf{V}_{3}$ in (\ref{HMN-background-K-asymp-2}), and
$\mathbf{V}_{i}$ in (\ref{HMN-background-K-asymp-3}) degenerate to
matrices with linear dependent columns. In conclusion, the method
with naive fluctuation symbol extensions is unsuccessful to create
channel randomness in constant channels to apply Cadambe-Jafar
scheme to achieve interference alignment.

While for slow changing channels as in the same expression of
(\ref{HMN-diag-H-h-single-extension-slowly-changing}), all the
statuses has a relationship as in
(\ref{HMN-diag-H-h-slow-changing}). It needs further validation and
proof whether IA scheme is applicable.

\subsection{Novel Design: Double-Layered Symbol Extensions}

The naive fluctuation coding to construct time-variant channels in
(\ref{HMN-alpha-h-beta}) and (\ref{HMN-diag-H-h-single-extension})
is proved to be not successful according to
(\ref{HMN-background-T-matrix-single-extension}). However, it still
inspires important clues for a novel achievable design. This novel
method is called double-layered symbol extensions.

For the sake of simplicity, let $T$ be an even number. Divide all
$T$ time slots into two pieces, i.e. from $1$ to $T/2$ as the first
piece, and from $(T/2+1)$ to $T$ as the second piece. Pair each two
time slots in each piece respectively in sequence, i.e. $1$ and
$(T/2+1)$, and $2$ and $(T/2+2)$ etc. It is equivalent to add two
statuses of channels to form one virtual status of the network. So
there are totally $T/2$ statuses, and the $(t)$-th virtual channel
status is obtained from the $t$-th time slot and $(T/2+t)$-th time
slot. In particular, $\mathrm{\tilde h}_{kj}^{t}$ is constructed as
following: \normalsize
\begin{equation}\label{HMN-alpha-h-beta-double-extension}
\begin{aligned}
\mathrm{\tilde
h}_{kj}^{[{t}]}=\beta_k^{[t]}\mathrm{h}_{kj}^{[t]}\alpha_j^{[t]}+\beta_k^{[T/2+t]}\mathrm{h}_{kj}^{[T/2+t]}\alpha_j^{[T/2+t]}
\end{aligned}
\end{equation}
\normalsize

To implement the design of (\ref{HMN-background-K-asymp-1}),
(\ref{HMN-background-K-asymp-2}) and
(\ref{HMN-background-K-asymp-3}), the length of symbol extensions is
set as $T=2[{(n+1)}^N+n^N]$ where $n\in\mathbb{N}$ and
$N=(K-1)(K-2)-1$. Then the time-extended effective channel
$\mathbf{H}_{kj}$ in (\ref{HMN-diag-H-h-single-extension}) is
updated and replaced by the following equation:
\begin{equation}\label{HMN-diag-H-h-double-extension}
\begin{aligned}
&\mathbf{H}_{kj}=\text{Diag}\{\mathrm{\tilde h}_{kj}^{[1]},
\mathrm{\tilde h}_{kj}^{[2]}, \cdots, \mathrm{\tilde
h}_{kj}^{[T/2]}\}\\
&=\text{Diag}\{\beta_k^{[1]}\mathrm{h}_{kj}^{[1]}\alpha_j^{[1]}+\beta_k^{[T/2+1]}\mathrm{h}_{kj}^{[T/2+1]}\alpha_j^{[T/2+1]},
\beta_k^{[2]}\mathrm{h}_{kj}^{[2]}\alpha_j^{[2]}+\beta_k^{[T/2+2]}\mathrm{h}_{kj}^{[T/2+2]}\alpha_j^{[T/2+2]}, \cdots\\
&\hspace{50mm}, \beta_k^{[T/2]}\mathrm{h}_{kj}^{[T/2]}\alpha_j^{[T/2]}+\beta_k^{[T]}\mathrm{h}_{kj}^{[T]}\alpha_j^{[T]}\}\\
\end{aligned}
\end{equation}

Observe that $\mathbf{H}_{kj}$ in
(\ref{HMN-diag-H-h-double-extension}) is effectively time-variant
and has the new dimension $\mathbf{H}_{kj}\in\mathbb{C}^{(T/2)\times
(T/2)}$. It could be further decomposed into the following equation:
\small
\begin{equation}\label{HMN-diag-H-h-double-extension-decompose}
\begin{aligned}
&\hspace{0mm}\mathbf{H}_{kj}=\Xi_k^{\star}\Delta_{kj}^{\star}\Omega_j^{\star}+\cdot\Xi_k^{\circ}\Delta_{kj}^{\circ}\Omega_j^{\circ}\\
&\Xi_k^{\star}=\text{Diag}\{\beta_k^{[1]}, \beta_k^{[2]}, \cdots,
\beta_k^{[T/2]}\}, \Xi_k^{\circ}=\text{Diag}\{\beta_k^{[T/2+1]},
\beta_k^{[T/2+2]}, \cdots, \beta_k^{[T]}\}\\
&\Delta_{kj}^{\star}=\text{Diag}\{\mathrm{h}_{kj}^{[1]},
\mathrm{h}_{kj}^{[2]}, \cdots, \mathrm{h}_{kj}^{[T/2]}\},
\Delta_{kj}^{\circ}=\text{Diag}\{\mathrm{h}_{kj}^{[T/2+1]},
\mathrm{h}_{kj}^{[T/2+2]}, \cdots, \mathrm{h}_{kj}^{[T]}\}\\
&\Omega_j^{\star}=\text{Diag}\{\alpha_j^{[1]}, \alpha_j^{[2]},
\cdots, \alpha_j^{[T/2]}\},
\Omega_j^{\circ}=\text{Diag}\{\alpha_j^{[T/2+1]},
\alpha_j^{[T/2+2]},
\cdots, \alpha_j^{[T]}\}\\
&\Xi_k^{\star},\ \Xi_k^{\circ},\ \Delta_{kj}^{\star},\
\Delta_{kj}^{\circ},\ \Omega_j^{\star},\
\Omega_j^{\circ}\in\mathbb{C}^{(T/2)\times (T/2)}
\end{aligned}
\end{equation}
\normalsize

Then check the intermediate matrix $\mathbf{T}_{kl}$ in
(\ref{HMN-background-T-matrix}) again as following: \small
\begin{equation}\label{HMN-background-T-matrix-double-extension}
\begin{aligned}
&\mathbf{T}_{kl}=\mathbf{H}_{21}{(\mathbf{H}_{23})}^{-1}\mathbf{H}_{13}{(\mathbf{H}_{k1})}^{-1}\mathbf{H}_{kl}{(\mathbf{H}_{1l})}^{-1}\\
&=
(\Xi_2^{\star}\Delta_{21}^{\star}\Omega_1^{\star}+\Xi_2^{\circ}\Delta_{21}^{\circ}\Omega_1^{\circ})\\
&\ \ \ \
(\Xi_2^{\star}\Delta_{23}^{\star}\Omega_3^{\star}+\Xi_2^{\circ}\Delta_{23}^{\circ}\Omega_3^{\circ})^{-1}
(\Xi_1^{\star}\Delta_{13}^{\star}\Omega_3^{\star}+\Xi_1^{\circ}\Delta_{13}^{\circ}\Omega_3^{\circ})
(\Xi_k^{\star}\Delta_{k1}^{\star}\Omega_1^{\star}+\Xi_k^{\circ}\Delta_{k1}^{\circ}\Omega_1^{\circ})^{-1}\\
&\ \ \ \
(\Xi_k^{\star}\Delta_{kl}^{\star}\Omega_l^{\star}+\Xi_k^{\circ}\Delta_{kl}^{\circ}\Omega_l^{\circ})
(\Xi_1^{\star}\Delta_{1l}^{\star}\Omega_l^{\star}+\Xi_1^{\circ}\Delta_{1l}^{\circ}\Omega_l^{\circ})^{-1}
\end{aligned}
\end{equation}
\normalsize

Based on the key theorem in Cadambe-Jafar scheme \cite[Theorem
1]{IA-DOF-Kuser-Interference}, we look into the $K$-pair
single-antenna network in constant channels as in
(\ref{HMN-diag-H-h-constant}) and
(\ref{HMN-diag-H-h-single-extension-decompose}), and slowly changing
channels as in (\ref{HMN-diag-H-h-slow-changing}) and
(\ref{HMN-diag-H-h-single-extension-slowly-changing-decompose}). If
effective channels are constructed with double-layered symbol
extensions as in (\ref{HMN-diag-H-h-double-extension}) and
(\ref{HMN-diag-H-h-double-extension-decompose}), then conventional
Cadambe-Jafar scheme in (\ref{HMN-background-K-asymp-1}),
(\ref{HMN-background-K-asymp-2}) and
(\ref{HMN-background-K-asymp-3}) could be much likely to be applied
on the effective channels to approach $K/4$ DoF for the network.
The detail achievable scheme is provided in \textit{Theorem 1} in
\cite[Theorem 1]{IA-DOF-Kuser-Interference}, in which \cite[Section
IV, Subsection B]{IA-DOF-Kuser-Interference} dealt with the 3-user
case and then \cite[Appendix III]{IA-DOF-Kuser-Interference} coped
with the arbitrary $K$-user case.

First, consider the alignment/overlapping condition of interference
subspaces of (\ref{HMN-background-conditionKuserI}). It is easily
verified that the design of precoders in
(\ref{HMN-background-K-asymp-1}), (\ref{HMN-background-K-asymp-2})
and (\ref{HMN-background-K-asymp-3}) satisfy the alignment
condition. It does not require any special features of
$\mathbf{H}_{kj}$ or $\mathbf{T}_{kl}$, so that the double-layered
symbol extensions  do not impact the IA scheme in terms of
(\ref{HMN-diag-H-h-double-extension-decompose}) and
(\ref{HMN-background-T-matrix-double-extension}).

Second, it is necessary to verify that the desired signals are
composed of linearly independent streams and at the same time they
are linearly independent of the interferences so that the streams
could be decoded by zero-forcing the interference.

Without losing generality, only take the received signal vectors at
the 1-st receiver as an example:
$\mathbf{R}=[\mathbf{H}_{11}\mathbf{V}_{1}\
\mathbf{H}_{12}\mathbf{V}_{2}]$. Notice the dimension is set as
$\mathbf{R}\in\mathbb{C}^{(T/2)\times (T/2)}$. As mentioned,
$\mathbf{H}_{12}\mathbf{V}_{2}$ represents all the aligned
interference subspaces from different transmitters to the 1-st
receiver. Therefore, in order to prove
$\mathbf{H}_{11}\mathbf{V}_{1}$ has full linearly independent
columns, it only needs to show $\mathbf{R}$ has full linearly
independent columns, i.e. the matrix $\mathbf{R}$ has full rank of
$T/2$.

Transform $\mathbf{R}$ to an equivalent matrix
$\mathbf{S}=[\mathbf{V}_{1}\
(\mathbf{H}_{11})^{-1}\mathbf{H}_{12}\mathbf{V}_{2}]$. In detail, it
is composed of $(n+1)^N$ columns in the form of
$\prod\mathbf{T}_{kl}^{n_{kl}}\mathbf{w}$ and $n^N$ columns in the
form of
$(\mathbf{H}_{11})^{-1}\mathbf{H}_{12}\prod\mathbf{T}_{kl}^{n_{kl}}\mathbf{w}$.
Let the diagonal entries of $\mathbf{T}_{kl}$ be
$\lambda_{kl}^{\langle1\rangle}, \lambda_{kl}^{\langle2\rangle},
\ldots, \lambda_{kl}^{\langle T/2\rangle}$ and the diagonal entries
of $(\mathbf{H}_{11})^{-1}\mathbf{H}_{12}$ be
$\kappa^{\langle1\rangle}, \kappa^{\langle2\rangle}, \ldots,
\kappa^{\langle T/2\rangle}$. So that the $q$-th entry is obtained
from $q$-th and $(T/2+q)$-th time slots. Let $s=q$, $t=T/2+q$, then
$\lambda_{kl}^{\langle q\rangle}$ and $\kappa^{\langle q\rangle}$
are presented as:

\small
\begin{equation}\label{HMN-background-T-matrix-double-extension-element}
\begin{aligned}
&\lambda_{kl}^{\langle
q\rangle}=\\
&\frac{(\beta_2^{[s]}\mathrm{h}_{21}^{[s]}\alpha_1^{[s]}+\beta_2^{[t]}\mathrm{h}_{21}^{[t]}\alpha_1^{[t]})(\beta_1^{[s]}\mathrm{h}_{13}^{[s]}\alpha_3^{[s]}+\beta_1^{[t]}\mathrm{h}_{13}^{[t]}\alpha_3^{[t]})(\beta_k^{[s]}\mathrm{h}_{kl}^{[s]}\alpha_l^{[s]}+\beta_k^{[t]}\mathrm{h}_{kl}^{[t]}\alpha_l^{[t]})}
{(\beta_2^{[s]}\mathrm{h}_{23}^{[s]}\alpha_3^{[s]}+\beta_2^{[t]}\mathrm{h}_{23}^{[t]}\alpha_3^{[t]})(\beta_k^{[s]}\mathrm{h}_{k1}^{[s]}\alpha_1^{[s]}+\beta_k^{[t]}\mathrm{h}_{k1}^{[t]}\alpha_1^{[t]})(\beta_1^{[s]}\mathrm{h}_{1l}^{[s]}\alpha_l^{[s]}+\beta_1^{[t]}\mathrm{h}_{1l}^{[t]}\alpha_l^{[t]})}\\
\end{aligned}
\end{equation}
\normalsize

\small
\begin{equation}\label{HMN-background-T-matrix-double-extension-element-kappa}
\begin{aligned}
\hspace{-25mm}\kappa^{\langle q\rangle}=\frac{(\beta_1^{[s]}\mathrm{
h}_{12}^{[s]}\alpha_2^{[s]}+\beta_1^{[t]}\mathrm{h}_{12}^{[t]}\alpha_2^{[t]})}
{(\beta_1^{[s]}\mathrm{h}_{11}^{[s]}\alpha_1^{[s]}+\beta_1^{[t]}\mathrm{h}_{11}^{[t]}\alpha_1^{[t]})}\\
\end{aligned}
\end{equation}
\normalsize

Then the matrix $\mathbf{S}$ is composed of elements of
$\lambda_{kl}^{\langle q\rangle}$ and $\kappa^{\langle q\rangle}$.
As mentioned in the beginning, the scheme is based on the same
procedure in \cite[Section IV, Subsection
B]{IA-DOF-Kuser-Interference} and \cite[Appendix
III]{IA-DOF-Kuser-Interference}. The detail is not repeatedly
described here.

To apply the achievable scheme to our case of either constant
channel in (\ref{HMN-diag-H-h-constant}) or slowly changing channel
in (\ref{HMN-diag-H-h-slow-changing}), two fundamental conditions
are required. 1) notice a key requirement is that $\kappa^{\langle
q\rangle}$ is a random variable drawn from a continuous distribution
so that it has probability zero to take a value of the corresponding
linear equations. 2) Furthermore, notice another key requirement
that all $\lambda_{kl}^{\langle q\rangle}$ are drawn independently
from a continuous distribution and they are all distinct almost
surely so that they have probability zero to be equal to the roots
of corresponding finite degree polynomials.

For our case, check
(\ref{HMN-background-T-matrix-double-extension-element}) and
(\ref{HMN-background-T-matrix-double-extension-element-kappa}) which
guarantee that $\lambda_{kl}^{\langle q\rangle}$ and
$\kappa^{\langle q\rangle}$ are random values since the gains
$\alpha_i^{[t]}$ and $\beta_j^{[t]}$ are randomly generated from
continuous distributions. It is also obvious that they are
independent because each distinct $q$-th entry only uses variables
within the corresponding two time slots. Finally, check
$\lambda_{kl}^{\langle q_1\rangle}$ and $\lambda_{kl}^{\langle
q_2\rangle}$ for $q_1\neq q_2$, and it is obvious they are distinct
so that it prevents the failure of
(\ref{HMN-background-T-matrix-single-extension}) as in the case of
naive fluctuation coding.
On condition the above randomness of $\lambda_{kl}^{\langle
q\rangle}$ and $\kappa^{\langle q\rangle}$ is guaranteed, it is
possible to proceed the IA scheme. However, it still needs further
validation and rigorous proof to check linear independency of all
high-rank exponentials in the constructed signal space.

\section{Primitive Numerical Results}

\begin{figure}[htpb]
  \centering
    \includegraphics[width=4.5in]{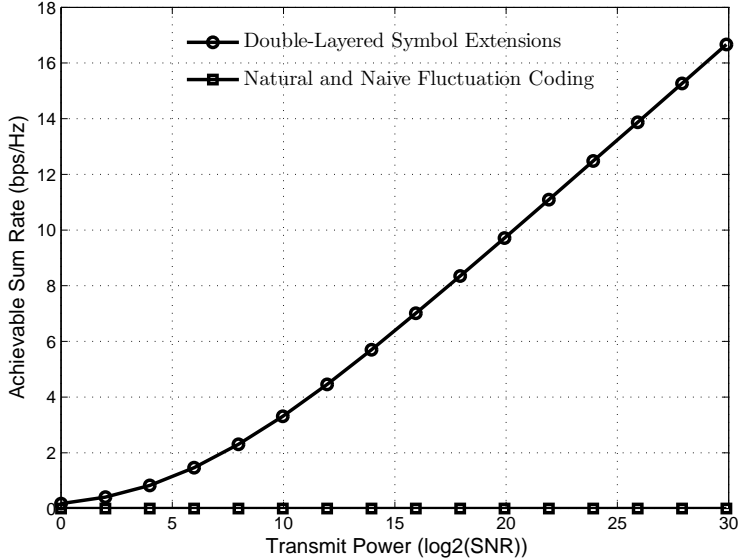}
  \caption{Sum Rate of a 3-User Network Applying IA with Symbol Extensions in Constant Channels}\label{HMN-3u-fig}
\end{figure}

To further validate the proposed novel method of double-layered
symbol extensions, numerical results are given as well. First, we
look at a 3-user network applying Cadambe-Jafar scheme of
interference alignment as in (\ref{HMN-background-K-asymp-1}),
(\ref{HMN-background-K-asymp-2}), and
(\ref{HMN-background-K-asymp-3}). For comparison, on one hand, we
use the natural and naive fluctuation coding as in
(\ref{HMN-alpha-h-beta}) and (\ref{HMN-diag-H-h-single-extension});
on the other hand we use the double-layered symbol extensions as in
(\ref{HMN-alpha-h-beta-double-extension}) and
(\ref{HMN-diag-H-h-double-extension}), both in constant channels.
These two cases are shown in Fig. \ref{HMN-3u-fig}.

In Fig. \ref{HMN-3u-fig}, since it is a 3-user network, i.e. $K=3$,
then the power component $N=(K-1)(K-2)-1=1$. For the natural and
naive fluctuation coding, set the length of symbol extensions
$T=2n+1$ and $n=2$, so that the network is supposed to obtain a
total DoF of $\frac{3n+1}{2n+1}=7/5$ if Cadambe-Jafar scheme works.
However, as shown in
(\ref{HMN-background-T-matrix-single-extension}), the network with
constant channels loses randomness by only applying the naive
fluctuation coding, so that normal transmission is not available as
shown in Fig. \ref{HMN-3u-fig}. For the double-layered symbol
extensions, set the length of symbol extensions $T=2(2n+1)$ and
$n=2$, so that the network is supposed to obtain a total DoF of
$\frac{3n+1}{2(2n+1)}=7/10$ with an additional $1/2$ factor due to
the double-layered extension. As shown in
(\ref{HMN-background-T-matrix-double-extension}), the network
successfully creates virtual time-variant channels to apply
effective interference alignment to approach $K/4$ DoF, so that it
is clearly shown in Fig. \ref{HMN-3u-fig} that the network obtains
$7/10$ DoF when $n=2$.

Theoretically, when $K=4$, the expected achievable DoF of the
network could approach $K/4=1$; when $K=5$, the expected achievable
DoF of the network could approach $K/4=1.25$. So that the achievable
DoF could surmount the previous obtained DoF of $1$ and $1.2$ in
\cite{multiplexing-gain-wireless} and
\cite{IA-asymmetric-complex-signaling-settling-nosratinia-conjecture-trans}
respectively. However, we are not able to illustrate the numerical
results due to limited computational capability. Set $K=5$ and
$N=(K-1)(K-2)-1=11$. When $n=81$, the total DoF is
$\frac{(n+1)^{11}+4n^{11}}{2[(n+1)^{11}+n^{11}]}=1.1995$; When
$n=82$, the total DoF is
$\frac{(n+1)^{11}+4n^{11}}{2[(n+1)^{11}+n^{11}]}=1.2001$. So in the
case of $n=82$, the DoF could surmount previous result of 1.2.
However, at this time, notice $(n+1)^{11}=1.2878e+021$,
$n^{11}=1.1271e+021$, and $(n+1)^{11}+n^{11}=2.4149e+021$, so that
$\mathbf{H}_{kj}\in\mathbb{C}^{(2.4149e+021)\times (2.4149e+021)}$,

$\mathbf{V}_{1}\in\mathbb{C}^{(2.4149e+021)\times (1.2878e+021)}$
and

$\mathbf{V}_{j}\in\mathbb{C}^{(2.4149e+021)\times (1.1271e+021)},
\forall j\in\mathcal{K}\backslash\{1\}$. The super large
dimensionality makes it difficult to shown numerical results.


\section{Conclusion}

In this work, we propose a novel method of double-layered symbol
extensions to generate virtual time-variant channels to apply
conventional Cadambe-Jafar scheme of interference alignment.

\bibliographystyle{IEEEtran}
\bibliography{Thesis}

\end{document}